\documentclass[11pt,superscriptaddress,aps,prd,preprint]{revtex4}
\usepackage[dvips]{graphicx}
\usepackage{amsfonts}
\usepackage{slashed}
\usepackage{amsmath}

\newcommand{\incps}[5]{\includegraphics[#2,#3][#4,#5]{#1}}
\newcommand{\incpicwh}[3]{\includegraphics[width=#2,height=#3]{#1}}
\newcommand{\pad}{\hspace{-0.1cm}}
\newcommand{\fgl}[1]{\hspace{0.4cm}#1\hspace{-0.4cm}}
\newcommand{\qedpic}[1]{\incpicwh{#1}{3.2cm}{1.8cm}}

\begin{document}

\title{Radiatively induced Lorentz-violating operator of mass dimension five in QED}

\author{T. Mariz}
\affiliation{Instituto de F\'\i sica, Universidade Federal de Alagoas, 57072-270, Macei\'o, Alagoas, Brazil}
\email{tmariz@if.ufal.br}

\date{\today}

\begin{abstract}
The first higher derivative term of the photon sector of Lorentz-violating QED, with operator of mass dimension $d=5$, is radiatively induced from the fermion sector, in which contains a derivative term with the dimensionless coefficient $g^{\lambda\mu\nu}$. The calculation is performed perturbatively in the coefficient for Lorentz violation, and due to the fact that the contributions are quadratically divergent, we adopt dimensional regularization. 
\end{abstract}

\maketitle

\section{Introduction}

The standard-model extension is a effective field theory that describes violations of Lorentz symmetry at attainable energies \cite{Kos,Kos2,Kos3}. It contains renormalizable terms formed by the contraction of Lorentz-violating operators of mass dimension $d=3$ and $d=4$ with coefficients of mass dimension $d=1$ and dimensionless, respectively. Recently studies of the Lorentz-violating quantum electrodynamics (QED) with operators of arbitrary mass dimensions have been carried out for the effective photon propagator, which include nonrenormalizable $CPT$-odd and $CPT$-even terms with operators of mass dimension $d\ge5$ (higher derivative extensions), contracted with coefficients of mass dimension $d\le-1$~\cite{Kos4}.

The first term with dimension five operator considered in Ref.~\cite{Kos4} is the first higher derivative extension of $CPT$-odd term given by 
\begin{equation}\label{hdt}
{\cal L}_k=\frac12 \epsilon^{\kappa\lambda\mu\nu}A_\lambda {(k_{AF})_\kappa}^{\alpha\beta}\partial_\alpha\partial_\beta F_{\mu\nu},
\end{equation}
with ${(k_{AF})_\kappa}^{\alpha\beta}=\frac1{3!}\epsilon_{\kappa\lambda\mu\nu}{\cal K}^{\lambda\mu\nu\alpha\beta}$, so that it can be rewritten as
\begin{equation}\label{hdt2}
{\cal L}_k={\cal K}^{\mu\nu\rho\alpha\beta} A_\mu \partial_\rho\partial_\alpha\partial_\beta A_\nu,
\end{equation}
where the coefficient ${\cal K}^{\mu\nu\rho\alpha\beta}$ has antisymmetry in the first three indices and symmetry in the last two.

In this work we are interested in studying whether the above term (\ref{hdt2}) can be induced through radiative corrections from the fermion sector of the Lorentz-violating QED, in the same way that was argued for the term with dimension four operator, the Chern-Simons term~\cite{Jac}. For other details and related references, see also~\cite{Col,Chu,Che,Per,Bon,Cha,Cer,Ada,Ada2,And,Mar,Alt,Gom}.

To study this issue, we will consider the derivative term with dimension four operator of the Lorentz-violating QED, which has the same $C$, $P$, and $T$ transformation properties as the higher derivative term (\ref{hdt2}), namely, the term with the coefficient $g^{\lambda\mu\nu}$ (see table I in Ref.~\cite{Kos5}). 

As the coefficient of the higher derivative term (\ref{hdt2}) has mass dimension $d=-1$, i.e. it is suppressed by one power of mass dimension, which may be related to the Planck mass $M_\mathrm{Pl}$ (so that ${\cal K}\sim M^{-1}_\mathrm{Pl}$) \cite{Mye}, such radiative induction could set small constraints on the dimensionless coefficient $g^{\lambda\mu\nu}$ of the Lorentz-violating QED. 

In the Ref.~\cite{Gro}, the smallness of the coefficients associated to the operators of mass dimension $d=3$ is explained as being due to a transmutation of dimension five operators into dimension three operators, via supersymmetry breaking. In our case, we have a dimension five operator being radiatively induced from a dimension four operator. However, it has been argued in~\cite{Kos2} that the coefficient $g^{\lambda\mu\nu}$ is already expected to be naturally suppressed, since it would be generate from higher dimensional operators.     

In next Sec.~\ref{rc}, we perform the radiative corrections of the higher derivative term (\ref{hdt2}) from the derivative term with the coefficient $g^{\lambda\mu\nu}$. In Sec.~\ref{drs} we analyze the dispersion relation, in order to observe the stability of the theory, and in Sec.~\ref{ne} we obtain numerical estimations for the Lorentz-violating coefficient. A summary is presented in Sec.~\ref{su}.

\section{Radiative corrections}\label{rc}

As we pointed out in the Introduction, the derivative term with the dimensionless coefficient $g^{\lambda\mu\nu}$ is a potential candidate to generate the higher derivative term (\ref{hdt2}), which has the coefficient ${\cal K}^{\mu\nu\rho\alpha\beta}$ of mass dimension $d=-1$. Note that for ${\cal K}\propto g$, we must have a certain linear combination of $g^{\lambda\mu\nu}$ in order to take into account the antisymmetry and symmetry properties of ${\cal K}^{\mu\nu\rho\alpha\beta}$. In fact, this is what happens, as we will see below.  
 
The fermion sector we are interested is described by the Lagrangian 
\begin{equation}
{\cal L}_f = \bar\psi\left(i\slashed{\partial}+\frac i2 g^{\kappa\lambda\mu}\sigma_{\kappa\lambda}\partial_\mu-m-\slashed{A}-\frac12 g^{\kappa\lambda\mu}\sigma_{\kappa\lambda}A_\mu\right)\psi,
\end{equation}
where $\sigma_{\kappa\lambda}=\frac i2 [\gamma_\kappa,\gamma_\lambda]$. The corresponding Feynman rules for the perturbative treatment are given as follows. The fermion propagator,
\begin{equation}
\raisebox{-0.4cm}{\incps{ferm.eps}{-1.5cm}{-.5cm}{1.0cm}{.5cm}}
 = \frac{i}{\slashed{p}-m}, 
\end{equation}
and the fermion-photon vertex,
\begin{equation}
\raisebox{-0.4cm}{\incps{vert.eps}{-1.5cm}{-.5cm}{1.0cm}{1.5cm}}
 = -i\gamma^\mu,
\end{equation}
are the usual ones. The coefficient for Lorentz violation lead to a insertion in the fermion propagator, 
\begin{equation}
\raisebox{-0.4cm}{\incps{fermgam.eps}{-1.5cm}{-.5cm}{1.0cm}{.5cm}}
 = \frac{i}{2}g^{\kappa\lambda\mu}\sigma_{\kappa\lambda}\,p_\mu, 
\end{equation}
and to a additional vertex, 
\begin{equation}
\raisebox{-0.4cm}{\incps{vertgam.eps}{-1.5cm}{-.5cm}{1.0cm}{1.5cm}}
 = -\frac{i}{2}g^{\kappa\lambda\mu}\sigma_{\kappa\lambda}.
\end{equation}

\begin{figure}
\centering
\begin{tabular}{cccc}
\fgl{(a)}\pad\qedpic{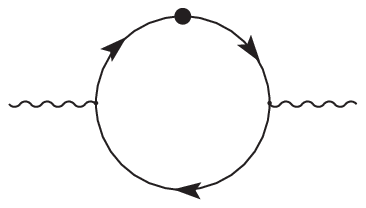}\pad &
\fgl{(b)}\pad\qedpic{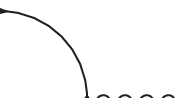}\pad &
\fgl{(c)}\pad\qedpic{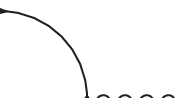}\pad &
\fgl{(d)}\pad\qedpic{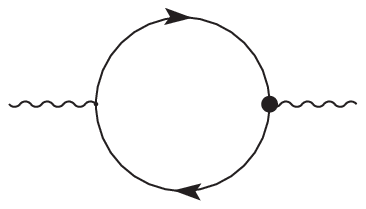}\pad 
\end{tabular}
\caption{One-loop contributions.}
\label{loops}
\end{figure}

The relevant one-loop contributions for the radiative corrections to the Lorentz-violating higher derivative term (\ref{hdt2}) are depicted in Fig.~\ref{loops}. These expressions are given by
\begin{subequations}\label{Pi}
\begin{eqnarray}
\label{Pia}i\Pi^{\mu\nu}_{(a)} &=& - \int\frac{d^4p}{(2\pi)^4} \mathrm{tr}\, (-i)\gamma^\mu iS(p)\frac i2 g^{\kappa\lambda\rho}\sigma_{\kappa\lambda}p_\rho iS(p)(-i)\gamma^\nu iS(p-k), \\
\label{Pib}i\Pi^{\mu\nu}_{(b)} &=& - \int\frac{d^4p}{(2\pi)^4} \mathrm{tr}\, (-i)\gamma^\mu iS(p)(-i)\gamma^\nu iS(p-k)\frac i2 g^{\kappa\lambda\rho}\sigma_{\kappa\lambda}(p_\rho-k_\rho) iS(p-k), \\
\label{Pic}i\Pi^{\mu\nu}_{(c)} &=& - \int\frac{d^4p}{(2\pi)^4} \mathrm{tr}\, \left(-\frac i2\right) g^{\kappa\lambda\mu}\sigma_{\kappa\lambda} iS(p)(-i)\gamma^\nu iS(p-k), \\
\label{Pid}i\Pi^{\mu\nu}_{(d)} &=& - \int\frac{d^4p}{(2\pi)^4} \mathrm{tr}\,  (-i)\gamma^\mu iS(p)\left(-\frac i2\right) g^{\kappa\lambda\nu}\sigma_{\kappa\lambda} iS(p-k),
\end{eqnarray}
\end{subequations}
with $S(p)=(\slashed{p}-m)^{-1}$ and $\mathrm{tr}$ means the trace over the Dirac matrices.  

In order to perform the above integrations, we first combine the denominators by employing Feynman parameter and later calculate the trace over Dirac matrices. Due to the fact that the integrals are quadratically divergent by power counting, we now adopt dimensional regularization by extending the spacetime from $4$ to $4-2\epsilon$ dimensions, so that  $d^4p/(2\pi)^4$ goes to $\mu^{2\epsilon}[d^{4-2\epsilon}p/(2\pi)^{4-2\epsilon}]$ and ${g^\alpha}_\alpha=4-2\epsilon$, in the conventional way. However, when we finally integrate over momenta and Feynman parameter the divergent contributions of Eqs.~(\ref{Pi}) cancel each other, as also observed in~\cite{Kos5}. The total result of the four expressions, $\Pi^{\mu\nu}_g=\Pi^{\mu\nu}_{(a)}+\Pi^{\mu\nu}_{(b)}+\Pi^{\mu\nu}_{(c)}+\Pi^{\mu\nu}_{(d)}$, is as follows
\begin{equation}\label{Pig}
\Pi^{\mu\nu}_g = -\frac{i m}{4\pi^2(k^2)^{3/2}}\left[\sqrt{k^2}-\frac{4m^2}{\sqrt{4m^2-k^2}}\arctan\left(\frac{\sqrt{k^2}}{\sqrt{4m^2-k^2}} \right)\right] {\cal G}^{\mu\nu\rho\alpha\beta} k_\rho k_\alpha k_\beta,
\end{equation}
where
\begin{equation}\label{G}
{\cal G}^{\mu\nu\rho\alpha\beta} = g^{\mu\nu\alpha}g^{\beta\rho}+g^{\mu\nu\beta}g^{\alpha\rho}-g^{\mu\rho\alpha}g^{\beta\nu}-g^{\mu\rho\beta}g^{\alpha\nu}-g^{\rho\nu\alpha}g^{\beta\mu}-g^{\rho\nu\beta}g^{\alpha\mu}.
\end{equation} 
Note that ${\cal G}^{\mu\nu\rho\alpha\beta}$ has the same antisymmetric and symmetric properties of ${\cal K}^{\mu\nu\rho\alpha\beta}$, as required for a consistent radiative correction of the higher derivative term (\ref{hdt2}).

The calculation of the trace over Dirac matrices in the expressions (\ref{Pi}) is not complicated in the $4-2\epsilon$ dimension, because we do not have the presence of the dimension-dependent $\gamma_5$ matrix. Therefore, we believe that there is no room for ambiguities in the above perturbative calculation.   

The Eq.~(\ref{Pig}) is well defined for $0<k^2<4m^2$, so that an extension for $k^2>4m^2$ requires an analytic continuation. However, as we are interested in a low-energy limit, in which $k^2<<4m^2$, this expression is enough for our purposes. In this way, the leading contribution, as $k^2/m^2$ tends to zero, is given by
\begin{equation}
\Pi^{\mu\nu}_g = \frac{i}{24m\pi^2}{\cal G}^{\mu\nu\rho\alpha\beta} k_\rho k_\alpha k_\beta + {\cal O}\left(\frac{k^2}{m^2}\right),
\end{equation}
which yields the Lagrangian 
\begin{equation}\label{Lg}
{\cal L}_g=\frac{1}{24m\pi^2}{\cal G}^{\mu\nu\rho\alpha\beta} A_\mu \partial_\rho\partial_\alpha\partial_\beta A_\nu,
\end{equation}
in coordinates space. This completes our radiative correction. 

Note that the totally antisymmetric part of $g^{\lambda\mu\nu}$ ($\tilde g^{\lambda\mu\nu}$), contained in Eq.~(\ref{Lg}), contributes to a higher derivative Chern-Simons term, namely, $\frac{1}{48m\pi^2}\,\epsilon^{\kappa\lambda\mu\nu} g_\kappa A_\lambda\Box F_{\mu\nu}$, with $g_\kappa=-\epsilon_{\kappa\lambda\mu\nu}\tilde g^{\lambda\mu\nu}$. Therefore, as the coefficients $g_\mu$ and $b_\mu$ always appear together, e.g. through a field redefinition \cite{Kos2,Col2}, we also expect a radiative contribution coming from $b_\mu$ for this higher derivative Chern-Simons extension. 

Another term that can be contained in (\ref{Lg}) is the term of the photon sector of the Myers-Pospelov model \cite{Mye}, given by $g\epsilon^{\kappa\mu\nu\rho}n_\kappa n^\alpha n^\beta A_\mu \partial_\alpha\partial_\beta\partial_\rho A_\nu$, which has been extensively studied in literature \cite{Mon,Bol,Mon2,JacLib,MarMon,Bol2,Gal,Rey,Rey2,Gub,Rey3}, but essentially for purely timelike $n_\kappa$. However, recently it was shown in \cite{Rey3} that for this timelike preferred direction, unitarity and causality are violated. On the other hand, for the purely spacelike case both stability and analyticity are preserved while microcausality is highly suppressed, which indicates the possibility of a consistent quantization of the theory.

Analyzing the components of ${\cal G}^{\mu\nu\rho\alpha\beta}$ (\ref{G}), we observe that ${\cal G}^{ijk00}$ is identically zero, i.e. the timelike Myers-Pospelov term is not radiatively induced. This is consistent with the analysis performed in \cite{Rey3}, since this preferred direction has nonunitarity evolution.

Now choosing only the components  ${\cal G}^{0ijkl}$, ${\cal G}^{i0jkl}$, and ${\cal G}^{ij0kl}$ to be nonzero, we have the radiative induction of a kind of spacelike Myers-Pospelov term, because some components are missing. In fact, this corresponds to take into account in our calculation only the component $g^{0ij}$. Note that, as the other components $g^{ij0}$, $g^{0i0}$, and $g^{ijk}$ induce terms containing higher time derivatives, which would spoil unitarity (for a recent discussion, see \cite{Ans,Ale,Pos}), they must be ruled out.

\section{Dispersion relation}\label{drs}

The dispersion relation that emerges from the Maxwell Lagrangian extended by the term (\ref{Lg}), is written as follows \cite{Kos4}:
\begin{equation}\label{dr}
k^4+4(\hat k_{AF})^2k^2-4(\hat k_{AF}\cdot k)^2=0,
\end{equation}
where
\begin{equation}\label{hatKaf}
(\hat k_{AF})_\kappa=\frac1{3!}\frac{1}{24m\pi^2}\epsilon_{\kappa\lambda\mu\nu}{\cal G}^{\lambda\mu\nu\alpha\beta}k_\alpha k_\beta.
\end{equation}
As can be easily seen, the purely timelike $(\hat k_{AF})_\kappa$ is composed only of components $g^{ij0}$ and $g^{ijk}$. Therefore, as we are interested only in the component $g^{0ij}$, we restrict ourselves to the spacelike preferred direction. In this case, we have $(\hat k_{AF})_0=0$ and 
\begin{equation}
(\hat k_{AF})_i=\frac12\frac{1}{24m\pi^2}\epsilon_{i0jk}{\cal G}^{0jklm}k_lk_m,
\end{equation}
so that $\hat k_{AF}\cdot k=0$ for any situation, i.e. $(\hat k_{AF})_i$ and $k_i$ are always perpendicular.

We now assume, for simplicity,  the case of $k_\mu=(k_0,0,0,k_3)$. With this choice, we obtain $(\hat k_{AF})_1=\frac{1}{12m\pi^2}g^{023}k_3k_3$, $(\hat k_{AF})_2=-\frac{1}{12m\pi^2}g^{013}k_3k_3$, and $(\hat k_{AF})_3=0$, so that the dispersion relation (\ref{dr}) becomes 
\begin{equation}
(k_0^2-k_3^2)^2-4g^2k_3^4(k_0^2-k_3^2)=0,
\end{equation}
where $g^2=\frac{1}{144m^2\pi^4}[(g^{023})^2+(g^{013})^2]$. The solutions are a usual dispersion relation, $k_0^2=k_3^2$, and an unusual one,  
\begin{equation}\label{un}
k_0^2=k_3^2(1+4g^2k_3^2).
\end{equation}

Analyzing the above unusual expression (\ref{un}), we observe that the energy is always real and becomes usual when $g$ goes to zero. Another characteristic found is that the solution is within the light cone. These are strong indications that unitarity is preserved in the theory. In fact, these are also characteristics found in the spacelike dispersion relation of the Myers-Pospelov model \cite{Rey3}.

To observe this more clearly, let us analyze the pole structure of the Feynman propagator. For this, we focus on the poles of the function    
\begin{equation}
K=\frac{k^2}{k^4+4(\hat k_{AF})^2k^2-4(\hat k_{AF}\cdot k)^2},
\end{equation}
already obtained previously for the Chern-Simons parameter $(k_{AF})_\kappa$ case \cite{Ada}, but that is the same for our case. Note that for the case of purely spacelike $(\hat k_{AF})_\kappa$, we have 
\begin{equation}
K=\frac1{k_0^2-k_i^2-4(\hat k_{AF})_i^2},
\end{equation}
and hence the propagator is well-behaved. 


\section{Numerical estimations}\label{ne}

Bounds on the coefficient ${\cal G}$ of Eq.~(\ref{Lg}) can be easily obtained from bounds on the coefficient ${\cal K}$, in systems related to the cosmic microwave background (CMB) polarization~\cite{Kos6} and astrophysical birefringence~\cite{Kos3} (see also table XV in Ref.~\cite{Kos7}). Thus, we roughly estimate ${\cal G}\sim10^{-21}$ and $\sim10^{-33}$, respectively, where we considered $m$ as being the electron mass. In this way, we get small estimates for the coefficient $g^{\lambda\mu\nu}$ from radiative corrections, compatible with the maximal attained sensitivities~\cite{Kos7}. 

\section{Summary}\label{su}

In this work we show that the higher derivative term (\ref{hdt2}), the first higher derivative extension of Lorentz-violating Chern-Simons term, is radiatively induced from the fermion sector of the minimal Lorentz-violating QED. The calculation was performed by taking into account the derivative term with the coefficient $g^{\lambda\mu\nu}$, which has the same C, P, and T transformation properties as the higher derivative term (\ref{hdt2}). Due to the absence of $\gamma_5$ matrix, we believe that the perturbative calculation adopted here, is free of ambiguities. The component $g^{0ij}$ induces a kind of spacelike Myers-Pospelov term, in which the resulting theory appears to be unitarity. Finally, we observe that such radiative correction seems to be compatible with the expected smallness of the dimensionless coefficient $g^{\lambda\mu\nu}$.

\vspace{.5cm}
{\bf Acknowledgements.} I would like to thank J. R. Nascimento for helpful discussions and V. Alan Kosteleck\'y for valuable comments. This work was supported by Conselho Nacional de Desenvolvimento Cient\'{\i}fico e Tecnol\'{o}gico (CNPq).

\end{document}